\documentclass[prd,amsmath,amssymb,groupedaddress,letterpaper,twocolumn]{revtex4}
\usepackage{amsmath,amssymb,graphicx,bm}
\newcommand{\cL}{{\cal L}}
\newcommand{\cM}{{\cal M}}

\newcommand{\cO}{{\cal O}}
\newcommand{\cR}{{\cal R}}

\newcommand{\nn}{\nonumber}
\newcommand{\p}{\partial}

\begin{document}

\title{Thermal gravity, black holes and cosmological entropy}

\author{Stephen~D.H.~Hsu} \email{hsu@duende.uoregon.edu}
\author{Brian~M.~Murray} \email{bmurray1@uoregon.edu}

\affiliation{Institute of Theoretical Science, University of Oregon,
  Eugene OR 94703-5203}

\begin{abstract}
Taking seriously the interpretation of black hole entropy as the 
logarithm of the number of microstates, we argue that thermal 
gravitons may undergo a phase transition to a kind of black hole 
condensate. The phase transition proceeds via nucleation of black 
holes at a rate governed by a saddlepoint configuration whose 
free energy is of order the inverse temperature in Planck units.
Whether the universe remains in a low entropy state as opposed
to the high entropy black hole condensate depends sensitively on
its thermal history. Our results may clarify an old observation of
Penrose regarding the very low entropy state of the universe.
\end{abstract}
%\pacs{}

\maketitle

\section{Introduction} \label{intro}

Years ago Penrose noticed that the universe must have begun in a very
low entropy state \cite{penrose}. By considering the entropy of black
holes, he argued that the current state of the universe has
significantly lower entropy than the maximum possible entropy
state. For example, while holding the number of baryons fixed one
could increase the total entropy tremendously by letting matter
collapse into black holes \cite{grentropy}.
Indeed, it seems that while the matter degrees of freedom were born
hot, i.e., in a maximum entropy thermal state, the gravitational
degrees of freedom were born in a very special low entropy
state. Interpreting entropy as the logarithm of phase space volume, a
low entropy state is an exponentially unlikely state and hence can
only result from fine-tuned initial conditions \cite{ergodic}.
Reasoning along these lines suggests that spacetimes with numerous
horizons, perhaps resembling a dense agglomeration of black holes,
occupy an exponentially larger fraction of gravitational phase space
than smooth spacetimes like the usual Friedmann-Robertson-Walker (FRW)
cosmologies.  For related discussions, see, e.g., \cite{carroll,wald}
and references contained therein.

One may ask whether special initial conditions at the Planck scale are
sufficient to produce the low entropy universe we see today. It might
be the case that interactions with thermal matter in the early
universe inevitably cause the gravitational degrees of freedom to
thermalize as well. Such a thermal state, assuming ergodicity of
gravity, would likely evolve to a configuration of much higher
entropy, and hence a cosmology very different from the one we observe.

Since black holes are our only hint at the highly entropic
configurations of gravity \cite{horizon}, they should play a prominent
role in the transition from low entropy to high entropy spacetimes. In
this paper we suggest a specific mechanism involving the nucleation of
black holes from a thermal graviton state. We note that the
corresponding nucleation rate from a thermal matter state is much
smaller, and probably irrelevant cosmologically. The mechanism we
describe provides a plausible means by which Penrose's ergodic
evolution could proceed. We examine whether the transition to a new,
highly entropic, phase of condensed black holes can occur in standard
big bang cosmology. The result depends sensitively on the thermal
history of the universe at early times. Moreover, the relevant
energy scales are all higher than the energy scale at which an
inflationary epoch is usually assumed to take place. Therefore, we are
considering a phase transition which only may take place before and
not after inflation. Presumably, the probability for a given patch to
inflate would be affected by whether or not that patch has undergone a
phase transition to the high entropy phase.

We should note that there is not a consensus on the issue of whether
gravity is ergodic nor on the interpretation of the gap between the
maximum allowed and the actual entropy in an FRW spacetime. Tipler
\cite{tipler} showed that under a reasonable set of assumptions,
closed universes are technically not ergodic, i.e., there is no
Poincar\'e recurrence. Moreover, Barrow \cite{barrow} has pointed out
that in a spacetime restricted to be FRW there is necessarily an
entropy gap, i.e., the entropy in thermal radiation is much less than
the entropy associated with a black hole of horizon size. However, the
phenomena we described in the previous paragraph, which are
investigated in this paper, are independent of these larger questions
about general relativity. That is, the mechanism by which black holes
are nucleated occurs on sub-horizon time and length scales. The
statistical approach we take below is justified by the presence of a
thermal bath of gravitons or other particles, whose existence is not
in dispute. In this sense we do not require any assumption of
ergodicity, {\it except in some small sub-horizon patch.}

In Sec. \ref{stat} we consider black hole nucleation in a system of
thermal gravitons and compare to a thermal system of matter. In Sec.
\ref{therm} we show that gravitons may thermalize in the early
universe even if they started out cold. We determine the conditions
necessary for a phase transition to a black hole condensate via
percolation in Sec. \ref{perc}. Finally, in Sec. \ref{disc} we relate
these results to Penrose's observation.  We use Planck units
throughout, i.e., $\hbar = c = G = k_{\rm B} = 1$.

\section{Statistical mechanics of gravitons} \label{stat}

Consider a box of hot gravitons. The probability for a fluctuation to
lead to a black hole of radius $R$ is
\begin{eqnarray}
  P(R) \sim N e^{-E/T}, \label{prob}
\end{eqnarray}
where $E=R/2$ is the energy of the black hole, and $N$ is the
multiplicity of microstates which, when coarse grained, appear as a
black hole of radius $R$.  The probability can be written as
\begin{eqnarray}
  P(R) \sim e^{-F/T}, \label{free_prob}
\end{eqnarray}
where $F = E - TS$ is the free energy, and $S = \ln N$. We assume that
black hole entropy is accounted for by gravitational microstates, as
suggested by results from string theory \cite{strominger}.  Using the
Bekenstein-Hawking formula \cite{bh_entropy} for black hole entropy
$S_{\rm BH} = A/4$, where $A=4 \pi R^2$ is the area, we see that
\begin{equation}
F(R) = R/2 - \pi T R^2. \label{free_energy}
\end{equation}
Strictly speaking, we want the free energy {\it relative} to that of
hot, flat space. This means we should subtract from
(\ref{free_energy}) a correction $F_0 (R) \sim - R^3 T^4$.  It is easy
to see that, near the saddlepoint found below, $F_0$ is a negligible
correction as long as the saddlepoint radius $R_*$ is much smaller
than the horizon size.

\begin{figure}
\includegraphics[width=8cm]{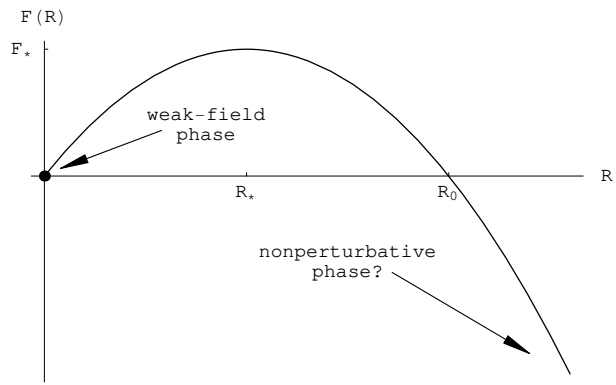}
\caption{Free energy versus black hole radius for a gas of hot
gravitons. Black holes of size $R < R_*$ shrink, leading to a
weak-field gravity phase. For $R > R_*$, however, black holes grow,
possibly giving rise to a new nonperturbative phase of gravity.}
\label{fig_energy}
\end{figure}

The radius that maximizes the free energy, the saddlepoint radius, is
given by $R_* = (4\pi T)^{-1}$.  We obtain $1 \ll R_* \ll H^{-1}$,
where $H^{-1} \sim 1/T^2$ is the horizon size for a radiation
dominated FRW universe, so our analysis encounters no difficulties
from quantum gravity or causality. The corresponding maximum free
energy is (see Fig. \ref{fig_energy})
\begin{eqnarray}
  F_* = (16 \pi T)^{-1}. \label{f_star}
\end{eqnarray}

At the saddlepoint radius the black hole temperature is just equal to
that of the heat bath.  Black holes with $0 < R < R_*$ shrink to zero
size, leaving a weak-field phase with a thermal population of graviton
states (gravity waves) on a smooth background metric. However, for $R
> R_*$, black holes grow without bound (they are colder than the
environment), and for $R > R_0 = (2\pi T)^{-1}$ the free energy is
actually negative, less than that of $R=0$. This instability may
indicate a new nonperturbative phase of gravity, which is not
asymptotically flat, and in which spacetime is all or partially filled
with black holes. Such a phase is highly entropic and occupies an
exponentially larger phase space volume than the smooth weak-field
phase.

The nucleation rate for supercritical black holes (which might be
thought of as bubbles of the new nonperturbative gravitational phase)
is controlled by the free energy at the saddlepoint, as in the usual
case of a first order phase transition (for early papers on nucleation
theory, see \cite{langer}). This yields
\begin{eqnarray}
  \lambda(T) \sim T^4 e^{-F_*/T}. \label{lambda}~
\end{eqnarray}
(Strictly speaking, the dimensional prefactor could be modified by
subleading terms in the exponent.) The physics is similar to that of
nonperturbative baryon number violation in the standard model. There,
the rate is controlled by the free energy of the electroweak
sphaleron, which is the saddlepoint configuration separating vacua of
different winding number \cite{sphaleron}.

Now consider a box of hot photons. If a fluctuation of size $R$ and
energy $E$ satisfies $E > R$ (we ignore factors of order one), it will
inevitably evolve into a black hole \cite{hoop}. Therefore,
Eqns. (\ref{prob}) and (\ref{free_prob}) for the probability for a
fluctuation to lead to a black hole of radius $R$ still hold. In this
case, however, the entropy is not proportional to $A$. In order to
evaluate the multiplicity $N$, we make use of a bound on the entropy
of a region of size $R$ filled with thermal radiation, originally
derived by 't Hooft \cite{'t Hooft}. By noting that matter in thermal
equilibrium, energy scales as $E \sim R^3 T^4$, and entropy as $S \sim
R^3 T^3$, and further requiring that the system not have already
undergone gravitational collapse, i.e., $E < R$, 't Hooft obtained the
bound $S < A^{3/4}$ (again, we ignore numerical factors).  Matter
configurations that lead to black holes saturate this bound.
Therefore, we see that the free energy of relativistic matter
configurations which evolve into black holes is of the form $F(R) \sim
R - T R^{3/2}$.

Once a fluctuation of sufficient size to lead to a black hole has
occurred, the evolution is then governed by the same physics as in the
original graviton case. We therefore compare the multiplicities of
configurations of photons and gravitons, $N_{\gamma}$ and $N_g$,
respectively, that will lead to black holes of critical size
$R_*$. For temperatures well below the Planck scale, $A \gg A^{3/4}$,
and so fluctuations of critical size in the photon gas are suppressed
relative to the case of a graviton gas by a factor of roughly
\begin{eqnarray}
  \frac{N_{\gamma}}{N_g} \sim 
    \exp\left(-\frac{1}{16 \pi T^2}\right).
\end{eqnarray}

The key difference between the two cases considered is the entropy
limit on thermal degrees of freedom. Ordinary matter cannot achieve
the entropic density of gravitational degrees of freedom, under the
assumption that black holes are coarse grained objects with $e^{A/4}$
gravitational microstates. Nucleation of black holes is much more
likely if these gravitational microstates are thermally occupied
(i.e., in a graviton heat bath), than if one starts with hot matter
and cold gravitons.

\section{Graviton thermalization} \label{therm}

Natural initial conditions for the universe might have both gravitons
and matter in thermal equilibrium. In some cases, however, thermal
matter leads to thermal graviton populations even when the gravitons
are initially cold. We can investigate the thermalization of gravitons
using a long-wavelength effective field theory (EFT) for quantum
gravity \cite{eft,donoghue}. If one is only interested in processes
occurring at energies sufficiently below the Planck scale, as is the
case here, there is no obstacle to using the standard EFT approach of
including all terms in the effective lagrangian that are consistent
with the symmetries of the system, in this case general coordinate
transformation invariance.

Without knowledge of the fundamental theory of quantum gravity, we are
not able to write down the Boltzmann equation for the evolution of the
phase space distribution for the gravitational degrees of freedom.
Boltzmann heuristics, however, motivate using $\Gamma(T) = H(T)$ as a
reasonable criterion for freeze-out of a given particle species, in
this case the graviton.  $\Gamma(T)$ is the interaction rate of
gravitons with the heat bath at temperature $T$, and $H(T)$ is the
Hubble expansion rate at temperature $T$. Below the decoupling
temperature $T_{\rm dec}$ the Hubble expansion rate is greater than
the interaction rate of gravitons with the heat bath, and gravitons
are decoupled.

Let us estimate the graviton decoupling temperature based on two
different scattering processes. First, consider the interaction $X_i
X_i \rightarrow gg$, where $X_i$ is any one of $N$ scalar, fermion or
vector particles, and $g$ is the graviton. In a relativistic gas,
Boltzmann heuristics suggest that the interaction rate is roughly
$\Gamma_{X_i X_i \rightarrow gg} \sim n_i \sigma_i$, where $n_i$ and
$\sigma_i$ are the number density and cross section, respectively, of
species $i$. At energies not too far below the Planck scale all
particle species are relativistic so $n_i \sim T^3$.  The matrix
element for this process goes as $T^2$ and the Hubble rate as $H \sim
N^{1/2}T^2$.  We see that the decoupling temperature obtained from
considering the process $\sum_i X_i X_i \rightarrow gg$ is $T_{\rm
dec} \sim N^{-3/2}$.  For a model with a large number of particle
species this scale may be well below the Planck scale.

This estimate, however, does not take into account interactions
between particle species. A similar argument based on considering the
process $X_i X_i \rightarrow \gamma g$, where $\gamma$ is another
particle species, leads to a decoupling temperature of roughly $T_{\rm
dec}^{\gamma} \sim N^{-1/2}\alpha^{-1}$, where $\alpha^{1/2}$ is the
coupling between $X_i$ and $\gamma$.  Again $T_{\rm dec}^{\gamma}$ may
be significantly below the Planck scale.

Let us obtain a more careful estimate of $T_{\rm dec}^{\gamma}$ in a
specific model: long-wavelength quantum gravity (EFT) coupled to
scalar QED (see, e.g., \cite{scalar_qed}). For a tree level
calculation, the relevant terms in the effective lagrangian are:
\begin{eqnarray}
  \cL &=&
    |g|^{\frac{1}{2}}\left[g^{\mu\nu}(D_{\mu}\phi)^*(D_{\nu}\phi)
    - \tfrac{1}{4}g^{\mu\nu}g^{\rho\sigma}
    F_{\mu\rho}F_{\nu\sigma} \right] \nn \\
    &+&\frac{1}{16 \pi}|g|^{\frac{1}{2}}\cR,
    \label{L_qed}
\end{eqnarray}
where $D_{\mu} = \p_{\mu} + ieA_{\mu}$, $F_{\mu\nu} = \p_{\mu}A_{\nu}
- \p_{\nu}A_{\mu}$, and $\cR$ is the scalar curvature.  The metric
$g_{\mu\nu}$ is expanded about the Minkowski metric $\eta_{\mu\nu} =
{\rm diag}(1,-1,-1,-1)$, i.e., $g_{\mu\nu}=\eta_{\mu\nu} + \sqrt{32
\pi} h_{\mu\nu}$, and Feynman rules are obtained by the standard
procedure. The cross section for the interaction $\phi\phi \rightarrow
\gamma g$ is found to be
\begin{eqnarray}
  \sigma_{\phi\phi \rightarrow \gamma g} = \frac{32 \pi}{6}\alpha,
    \label{sigma}
\end{eqnarray}
where $\alpha = e^2 / 4\pi$. Assuming local thermal equilibrium, the
scalar number density is $n_{\phi} = (2 \zeta(3)/
\pi^2)T^3$. Moreover, if there are $N_s$ species of scalar particles,
the graviton decoupling temperature is
\begin{eqnarray}
  T_{\rm dec}^{\gamma} \simeq \frac{3}{5\alpha N_s^{1/2}}.
    \label{T_dec}
\end{eqnarray}

For consistency we must also estimate the temperature below which we
can trust the above calculation, i.e., estimate the energy scale below
which the EFT description is valid. Let us look at the next order
contribution to the matrix element for this process:
\begin{eqnarray}
  \cM_{\phi\phi \rightarrow \gamma g} = 
    \cM_{\phi\phi \rightarrow \gamma g}^{(1)}
    + \cM_{\phi\phi \rightarrow \gamma g}^{(3)} + \ldots.
    \label{matrix_el}
\end{eqnarray}
Notice $\cM_{\phi\phi \rightarrow \gamma g}^{(1)} = \cO
(e\sqrt{32\pi}q)$, where $q$ is some typical energy scale of the
interaction. In order to obtain a conservative estimate of
$\cM_{\phi\phi \rightarrow \gamma g}^{(3)}$, we will assume that all
diagrams interfere constructively. Then $\cM_{\phi\phi \rightarrow
\gamma g}^{(3)} = \cO (FLe(\sqrt{32\pi}q)^3)$, where $F$ is the number
of graphs at this order and $L$ is a loop factor associated with each
graph. The EFT is valid for
\begin{eqnarray}
  q \ll q_{\rm pert} = (32 \pi FL)^{-1/2}.
    \label{q_pert}
\end{eqnarray}
Taking $F=27$ and $L = 1/4\pi^2$ we get $q_{\rm pert} \simeq 10^{-1}$.
By comparing Eqs. (\ref{T_dec}) and (\ref{q_pert}), we see that
$T_{\rm dec}^{\gamma} \ll q_{\rm pert}$ in the large $N_s$ limit, due
to the fact that $T_{\rm dec}^{\gamma} \sim N_s^{-1/2}$, while $q_{\rm
pert}$ is independent of $N_s$.  Thus, there is a class of models
(those with large numbers of matter fields) in which gravitons are
copiously produced and interact frequently, at energy scales where the
effective field theory applies.

One could perform a similar calculation with a more realistic model
for the matter content of the early universe.  At the energy scales of
interest, say $10^{16}$ GeV, the matter interactions may, for example,
be described by a grand unified theory (GUT) \cite{gut} with $N_s \sim
10^2$ and couplings not much smaller than unity.  A calculation in
such a model would likely yield the same conclusion, i.e., that there
was an epoch in which even initially cold gravitons interacted
strongly enough with the heat bath that they were themselves
thermalized.

\section{Percolation of black holes} \label{perc}

\begin{figure}
\includegraphics[width=8cm]{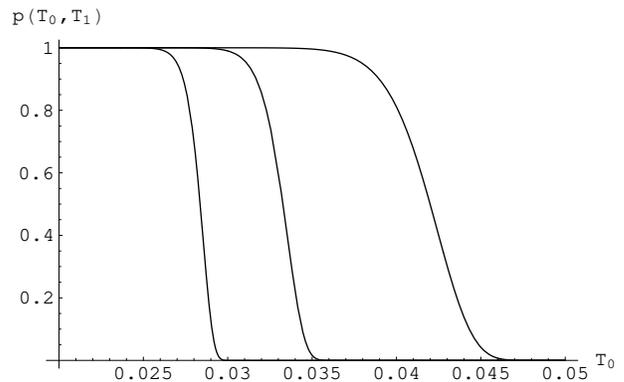}
\caption{Volume fraction in the weak-field phase of gravity as a
function of the temperature $T_0$ at which gravitons enter thermal
equilibrium. Shown are plots for three different values of the
expansion speed, $b = 10^{-1}$, $10^{-2}$, $10^{-3}$ (left to right).
Here the cut-off temperature $T_1 = 10^{-6}$, and the effective number
of degrees of freedom $g_* = 10^2$.}
\label{fig_perc}
\end{figure}

In light of the previous sections, it is interesting to consider the
possibility of a first order phase transition in an FRW universe with
a thermal population of gravitons.  Assuming that gravitons are in
thermal equilibrium in the early universe between an initial time
$t_0$ and a later time $t_1$, the volume fraction in the weak-field
gravity phase is given by:
\begin{eqnarray}
  p(t_0,t_1) = \exp\left[
    -\int_{t_0}^{t_1} dt' V(t',t_1)\lambda(t') \right].
    \label{vol_frac}
\end{eqnarray}
In flat FRW spacetime
\begin{eqnarray}
  V(t',t_1) = \frac{4\pi}{3}\left[a(t')b
    \int_{t'}^{t_1} \frac{dt''}{a(t'')}\right]^3.
\end{eqnarray}
Here $a$ is the scale factor, and $b$ is the (constant) speed at which
the black holes expand.  Eq. (\ref{vol_frac}) is a standard formula
from old inflationary cosmology \cite{guth}.  $b$ is governed by
relativistic physics and, therefore, should not be significantly
smaller than unity.  During a radiation dominated epoch the scale
factor goes as $a(t) \sim t^{1/2}$ and temperature and time are
related by $t \simeq 0.3 g_*^{-1/2}T^{-2}$. The volume fraction (now
as a function of temperature) is then of the form:
\begin{eqnarray}
  p(T_0,T_1) = \exp\left[
    -\frac{b^3}{g_*^2}f(T_0,T_1) \right].
\end{eqnarray}
We define the critical initial temperature $T_0^c$, so that if the
initial graviton temperature is larger than $T_0^c$, the volume
fraction in the weak-field phase is nearly zero. See
Fig. \ref{fig_perc} for a plot of $p(T_0,T_1)$ for $b = 10^{-1}$,
$10^{-2}$, and $10^{-3}$, with $T_1 = 10^{-6}$ and $g_* =
10^2$. Decreasing the expansion speed by two orders of magnitude has
the effect of increasing $T_0^c$ by less than a factor of
two. Increasing the number of effective degrees of freedom or the
cut-off temperature $T_1$ both have similar effects on $T_0^c$, i.e.,
they cause it to increase by a relatively small amount.

The critical initial temperature $T_0^c$ is of order $10^{-2}$ to
$10^{-1}$, which from Eq. (\ref{q_pert}) is the same order of
magnitude as the temperature at which quantum gravity becomes
perturbative.  Thus, taking the initial temperature of the big bang to
be roughly the scale at which gravity becomes perturbative, the
universe could avoid a phase transition to the nonperturbative black
hole phase, assuming it begins in the weak-field phase.

\section{Discussion} \label{disc}

We examined a possible first order phase transition of spacetime to a
black hole phase with high entropy.  Percolation of the high entropy
phase occurs if gravitons are ever in a thermal state with temperature
above $T_0^c$, either because they were born hot at the Planck epoch
or because they were thermalized due to interactions with thermal
matter. It seems possible, as suggested by entropic arguments, that
almost all of gravitational phase space is accounted for by the
nonperturbative phase. However, we find that the low entropy phase is
metastable over timescales which are exponentially sensitive to the
temperature, and potentially quite long.

One may wonder how inflation changes this conclusion. We note that
$T_0^c$ is higher than the energy scale at which inflation is usually
assumed to take place. If gravitons are never thermalized above a
temperature of $T_0^c$ then presumably inflation would simply take
place as originally envisioned. However, if gravitons are ever thermal
with a temperature above $T_0^c$ then we would speculate that it may
be less probable for a given patch to inflate, although depending on
the details of the model some non-zero probability may remain, even if
a phase transition to the high entropy black hole phase does occur.

Hot gravitons with temperature slightly below $T_0^c$ will not lead to
a phase transition; they will simply be red-shifted away. Both
gravitons and matter may be born hot, as long as the temperature of
the universe (either initially or after a period of inflation) is
never greater than $T_0^c$.  This does not require fine tuning because
$T_0^c$ is of the same order as $q_{\rm pert}$, the energy scale below
which quantum gravity effects are small. It may still be the case that
the initial conditions represent a subset of measure zero in the total
phase space, which is dominated by the nonperturbative black hole
phase \cite{anthropic}. However, our analysis does show that once the
initial choice of the low entropy phase is made, no transition to the
high entropy phase need occur.

These conclusions remain unchanged in a spacetime of arbitrary
dimension $d$. For hot gravitons in $d$ dimensions, the exponent
governing the nucleation rate of Eq. (\ref{lambda}) goes as $F_*/T
\sim T^{-(d-2)}$, while for matter $F_*/T \sim T^{-(d-1)(d-2)/2}$. For
$d > 3$, black hole nucleation is suppressed more strongly in the
matter system than in the gravitational one.  Moreover, as $d
\rightarrow \infty$, $T_0^c$ increases, meaning a transition to the
black hole condensate phase is less likely in a universe with a large
number of spacetime dimensions.

{\bf Note added}: After this paper was completed we became aware of
earlier work using Euclidean path integral methods in which
Eqs. (\ref{free_energy}), (\ref{f_star}), and (\ref{lambda}) were
independently derived \cite{gross}. In these calculations imaginary
time boundary conditions are applied to the gravitational
field. Therefore, those authors were also studying thermal gravity and
not only thermal matter. For discussion of black hole phase
transitions, see \cite{hu}.

\bigskip

\begin{center}
\textbf{Acknowledgments}
\end{center}

The authors thank R. Buniy and Y. S. Myung for useful comments. 
This work was supported by
the Department of Energy under DE-FG06-85ER40224.


\bigskip

\begin{thebibliography}{99}


\bibitem{penrose}
  R.~Penrose, {\it The Emperor's New Mind}, Oxford University Press, 
  Oxford, 1989;
  R.~Penrose, {\it The Road to Reality: A Complete Guide 
  to the Laws of the Universe}, Knopf, New York, 2005.

\bibitem{grentropy} Gravitational entropy is maximized by clumping
because gravity is a long range, unscreened, attractive force. This is
quite unlike non-gravitational systems (with screened or limited range
interactions) whose entropy is maximized when matter and energy are
uniformly distributed.


\bibitem{ergodic} The use of terms such as entropy, ergodicity or
phase space volume to describe configurations in general relativity
(spacetimes) might be unfamiliar.  However, classical general
relativity can be considered a dynamical system (i.e., a set of rules
for evolving a set of initial data given on some spacelike slice) like
any other. Then, if the usual assumptions of statistical mechanics
hold, time averaging is equivalent to ensemble averaging (ergodicity),
and the system is overwhelmingly likely to be found in states of
maximum entropy.

\bibitem{carroll}
  S.~M.~Carroll and J.~Chen,
  %``Spontaneous inflation and the origin of the arrow of time,''
  arXiv:hep-th/0410270;
  S.~M.~Carroll and J.~Chen,
  %``Does inflation provide natural initial conditions for the universe?,''
  arXiv:gr-qc/0505037;
  R.~Holman and L.~Mersini-Houghton,
  %``Why the universe started from a low entropy state,''
  arXiv:hep-th/0511102;
  R.~Holman and L.~Mersini-Houghton,
  %``A fly in the SOUP,''
  arXiv:hep-th/0511112.

\bibitem{wald}
  R.~M.~Wald,
  %``The Arrow of Time and the Initial Conditions of the Universe,''
  arXiv:gr-qc/0507094.

\bibitem{tipler}
  F.~Tipler,
  %''General relativity, thermodynamics, and the Poincar\acute{e} cycle'' 
  Nature {\bf 280}, 203 (1979).

\bibitem{barrow}
  J.~D.~Barrow,
  %``Entropic Principles,''
  New Astron.\  {\bf 4}, 333 (1999)
  [arXiv:astro-ph/9903225].

\bibitem{horizon}
One can also associate an entropy with the presence of a cosmological
horizon. Interestingly, when a black hole is added to de Sitter space
the total entropy associated with both the black hole and cosmological
horizons decreases; see, e.g., 
  D.~Klemm and L.~Vanzo,
  %``Aspects of quantum gravity in de Sitter spaces,''
  JCAP {\bf 0411}, 006 (2004)
  [arXiv:hep-th/0407255].
However, the interpretation of the entropy associated with a cosmological
horizon seems to us even more challenging than that of black holes, and so 
we choose to focus on the latter.

\bibitem{hoop}
  K.~S.~Thorne, {\it Nonspherical gravitational collapse:
  A short review}, in J.~R.~Klauder, {\it Magic Without
  Magic}, San Francisco 1972, 231-258;
  D.~M.~Eardley and S.~B.~Giddings,
  %``Classical black hole production in high-energy collisions,''
  Phys.\ Rev.\ D {\bf 66}, 044011 (2002)
  [arXiv:gr-qc/0201034];
  S.~D.~H.~Hsu,
  %``Quantum production of black holes,''
  Phys.\ Lett.\ B {\bf 555}, 92 (2003)
  [arXiv:hep-ph/0203154].

\bibitem{'t Hooft}
  G.~'t Hooft,
  %``Dimensional reduction in quantum gravity,''
  arXiv:gr-qc/9310026.

\bibitem{strominger}
  A.~Strominger and C.~Vafa,
  %``Microscopic Origin of the Bekenstein-Hawking Entropy,''
  Phys.\ Lett.\ B {\bf 379}, 99 (1996),
  [arXiv:hep-th/9601029].

\bibitem{bh_entropy}
  J.~D.~Bekenstein,
  %``Black Holes And Entropy,''
  Phys.\ Rev.\ D {\bf 7}, 2333 (1973);
  S.~W.~Hawking,
  %``Particle Creation By Black Holes,''
  Commun.\ Math.\ Phys.\  {\bf 43}, 199 (1975)
  [Erratum-ibid.\  {\bf 46}, 206 (1976)].

\bibitem{langer}
  J.~S.~Langer,
  %``Theory Of The Condensation Point,''
  Annals Phys.\  {\bf 41}, 108 (1967)
  [Annals Phys.\  {\bf 281}, 941 (2000)];
  J.~S.~Langer,
  %``Statistical Theory Of The Decay Of Metastable States,''
  Annals Phys.\  {\bf 54}, 258 (1969).

\bibitem{sphaleron}
  P.~Arnold and L.~D.~McLerran,
  %``Sphalerons, Small Fluctuations And Baryon Number Violation In Electroweak
  %Theory,''
  Phys.\ Rev.\ D {\bf 36}, 581 (1987);
  P.~Arnold and L.~D.~McLerran,
  %``The Sphaleron Strikes Back,''
  Phys.\ Rev.\ D {\bf 37}, 1020 (1988).

\bibitem{eft}
  S.~Weinberg,
  %``Phenomenological Lagrangians,''
  PhysicaA {\bf 96}, 327 (1979).

\bibitem{donoghue}
  J.~F.~Donoghue,
  %``General relativity as an effective field theory: The leading quantum
  %corrections,''
  Phys.\ Rev.\ D {\bf 50}, 3874 (1994),
  [arXiv:gr-qc/9405057].

\bibitem{scalar_qed}
  N.~E.~J.~Bjerrum-Bohr,
  %``Leading quantum gravitational corrections to scalar QED,''
  Phys.\ Rev.\ D {\bf 66}, 084023 (2002),
  [arXiv:hep-th/0206236].

\bibitem{gut}
  G.~G.~Ross, {\it Grand Unified Theories},
  Benjamin/Cummings, 1984.

\bibitem{guth}
  A.~H.~Guth and E.~J.~Weinberg,
  %``Could The Universe Have Recovered From A Slow First Order Phase
  %Transition?,''
  Nucl.\ Phys.\ B {\bf 212}, 321 (1983).

\bibitem{anthropic} Of course, one could always invoke anthropic
arguments to exclude the black hole phase, as it is hard to imagine
how life might evolve there.

\bibitem{gross}
  D.~J.~Gross, M.~J.~Perry and L.~G.~Yaffe,
  %``Instability Of Flat Space At Finite Temperature,''
  Phys.\ Rev.\ D {\bf 25}, 330 (1982);
  B.~F.~Whiting and J.~W.~York,
  %``Action Principle And Partition Function For The Gravitational Field In
  %Black Hole Topologies,''
  Phys.\ Rev.\ Lett.\  {\bf 61}, 1336 (1988).

\bibitem{hu}
  G.~J.~Stephens and B.~L.~Hu,
  %``Notes on black hole phase transitions,''
  Int.\ J.\ Theor.\ Phys.\  {\bf 40}, 2183 (2001)
  [arXiv:gr-qc/0102052].



\end{thebibliography}
\end{document}